\documentclass[%
 reprint,
 aip,
 amsmath,amssymb,
floatfix,
]{revtex4-1}
\usepackage{xcolor}
\usepackage{natbib}
\usepackage{blindtext}
\usepackage{float}
\usepackage{graphicx}
\usepackage{dcolumn}
\usepackage{bm}
\usepackage{soul}
\usepackage{url}
 \usepackage{hyperref}     
 
\begin{document}
\preprint{APS/123-QED}


\title{
Anharmonic effects on the reflectivity of CaS and MgS: A first-principles based study}
\author{Artem Chmeruk}
 \email{chmeruk@gfz-potsdam.de}
\author{Maribel N\'u\~nez-Valdez}
\email{mari\_nv@gfz-potsdam.de}
\affiliation{Deutsches GeoForschungsZentrum GFZ, Telegrafenberg, 14473 Potsdam, Germany}
\affiliation{Institut f\"ur Geowissenschaften, Goethe-Universit\"at Frankfurt, D-60438 Frankfurt am Main, Germany}
\date{\today}%
             %

\begin{abstract}
We employ systematic calculations based on density functional theory to model the reflectivity of CaS and MgS in the infrared region. 
We show that in addition to the modeling using the harmonic approximation, an accurate spectral description requires the inclusion of anharmonic effects. Due to their conceptual simplicity, CaS and MgS are excellent systems for the explicit consideration of the anharmonicity, which we include here using a perturbative approach up to three-phonon scattering processes, and the consideration of isotopic disorder. All physical quantities, such as Born effective charges and dielectric constant,  necessary for the calculation of the reflectivity within the Lorentz model are extracted from our first-principles computations. To validate our predicted optical and transversal modes, and reflectivity spectra, we compare them to available experimental results. 
We find that the overall agreement is good, which supports the importance of the inclusion of  anharmonic terms in the modeling of optical properties in the infrared region.
\end{abstract}
\maketitle
\section{\label{sec:level1}INTRODUCTION}
The alkaline-earth metal sulfides XS (X = Be, Mg, Ca, Sr, Ba) have recently attracted interest in science and
technology because of their remarkable physical properties and wide applications, ranging from catalysis to
microelectronics \cite{phasetransit}. In particular, CaS- and MgS-based compounds are known for their extensive use as highly efficient photoluminescents, cathodoluminescents, and X-ray phosphors \cite{photoluminescence1987,photoluminescence1991,o1998crystal,kasano1984cathodoluminescence}. Additionally, the alkaline-earth and transition metal sulfides have been investigated for a potential use as electrode materials for electrochemical energy storage \cite{pervez2022evaluation,he2014stable,zhang2004mineral,zhang2012unusual,tang2015morphology}.  Also, the optical response of MgS and CaS in the radio and infrared range is of high interest in the study of Mercury's surface composition as previous research and surveys of the planet speculate that it is rich in volatiles \cite{Mercury1,Mercury2}. Lastly, these monosulfides can also be found in meteorites \cite{avril2013raman,keil1989enstatite,keil2007occurrence,dibb2022spectral}, therefore their thorough understanding could help to better describe the evolution of our solar system.

There have been a few experimental and theoretical studies of the optical properties in the infrared region of CaS and MgS. From experimental investigations, data obtained from absorption spectra are used to determine various physical quantities such as reflectivity, emissivity \cite{kaneko1982optical,hofmeister2003absorption}, etc. On the other hand, theoretical studies based on density functional theory (DFT) within the local density approximation (LDA)\cite{ceperley1980ground} have only modeled the atomic dynamics of these systems in the limit of the harmonic approximation \cite{bayrakci,duman2006first}.
Unfortunately, the lack of neutron scattering studies on these sulfides makes impossible a direct comparison of the predicted phonon dispersions with their experimental counterparts. 
The only physical quantity directly comparable from experimental and modeling results is the splitting between longitudinal and transverse optical modes, i.e., the LO/TO splitting, which originates from the degeneracy elimination between the LO and TO phonons at the Brillouin-zone center \cite{calandrini2021limits}. In this regard, there is an overall reasonable agreement among reported measurements and predicted data (Table \ref{table1}). However, the harmonic approximation predicts a reflectivity with a step-like behavior and sharp edges at both ends of the spectrum\cite{cardona2005fundamentals}. This predicted reflectivity behavior is in drastic disagreement with experimental spectra, which appear with smeared edges, especially at the high-wavenumber tail. This inconsistency is a direct consequence of the simplifications made in the harmonic approximation and can be remedied by inclusion of anharmonic effects.

\begin{figure}
\includegraphics[width=1\linewidth]{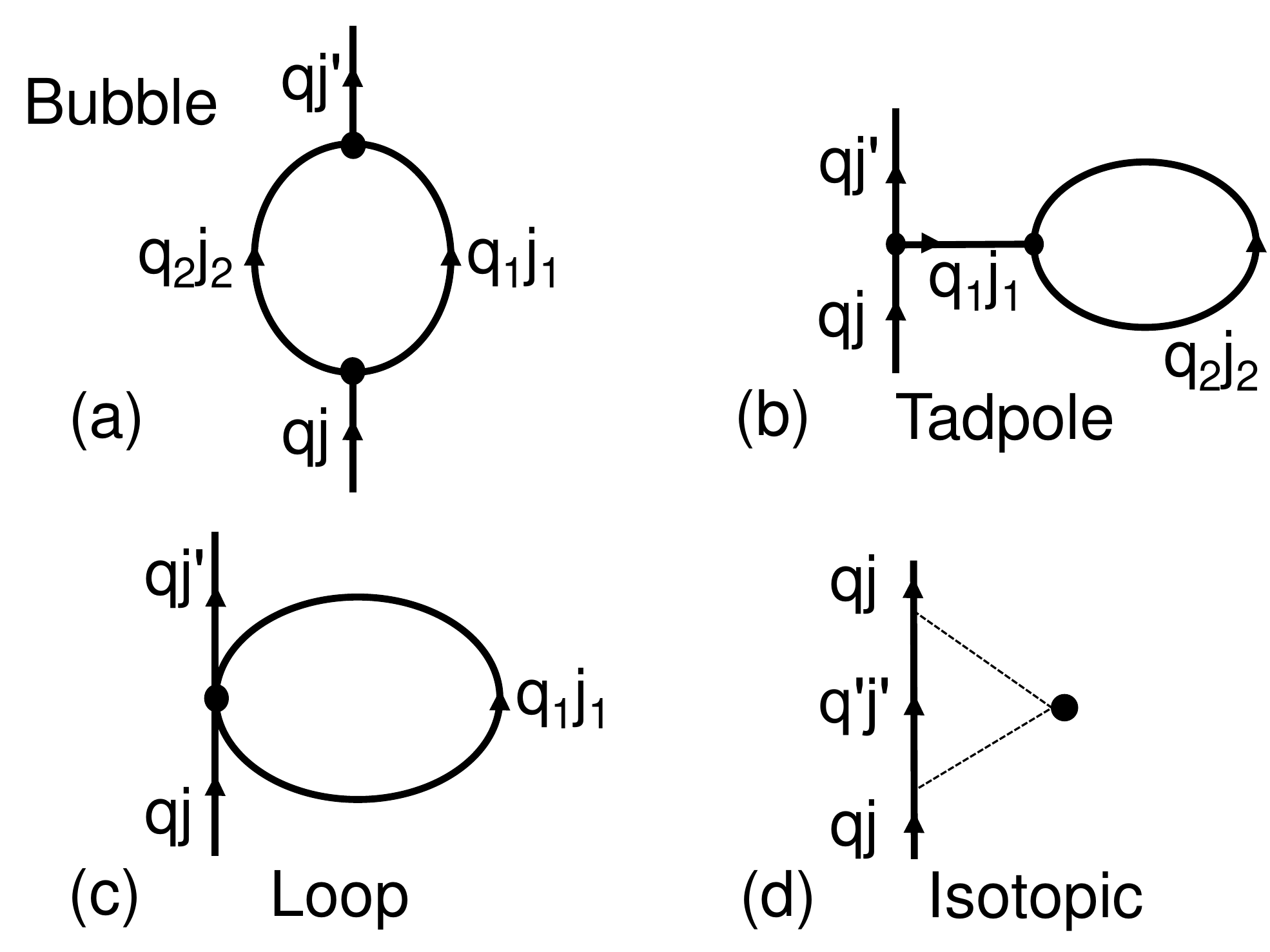}
\caption{\label{phon_diag}Diagrammatic anharmonic contributions considered in this work. Phonons are indexed by their wavevector $q$ and branch $j$. Incoming and outgoing phonons are represented by $qj$ and $qj'$, and internal phonon lines are given by $q_ij_i$ respectively \cite{maradudin1962scattering}. Diagrams (a) and (b) represent three-phonon scattering processes known as ``bubble" and ``tadpole" contributions. Diagram (c) shows a four-phonon scattering process  known as ``loop''. In our study, the only process entering in our model, as explained in the main text, is the bubble contribution. Diagram (d) represents isotope-disorder scattering \cite{tamura1983isotope}, the dashed lines represent phonon scatterings by the isotope shown here as the black dot. This process is also considered in our anharmonic modeling and it is obtained from second-order perturbation theory, using mass deviation as a perturbation.
}
\end{figure}

Thus, it has long been desired to include anharmonic effects in the first-principles atomic and molecular dynamics (MD) simulations. However, due to computational limitations, the explicit treatment of the phonon-phonon interactions has been limited, even though the analytical base has been available for a long time \cite{cowley1963lattice,maradudin1962scattering}. But in the last decade, the significant increase in computational power and efficiency has made possible to include anharmonic effects explicitly and on a rigorous basis. One way to consider directly anharmonic effects in a MD simulation is by coupling the system to a thermostat of a given temperature \cite{oganov2003all, hellman2011lattice}.  The consequent normal modes of motion obtained then contain the frequencies renormalized by the temperature, i.e., anharmonic interactions. Such an approach has been shown to successfully remedy the negative frequencies in certain elemental systems, persistent within the harmonic approximation \cite{hellman2011lattice}. The vibrational eigenfrequencies calculated in this way are ``dressed" by anharmonic phonon-phonon interactions up to, formally speaking, all orders. Therefore, in order to estimate individual contributions, a mapping scheme onto an effective model, containing all the contributions separately needs to be constructed \cite{klarbring2020anharmonicity}. Alternatively, one might wish to introduce anharmonicity in a progressive and more controlled way. This implementation is achieved in a perturbative manner based on the Cowley's expansion \cite{cowley1963lattice}. At the lowest order one obtains two three-phonon and one four-phonon contributions to the total energy as shown in Fig.~\ref{phon_diag}(a)-Fig.~\ref{phon_diag}(c). The main computational challenge to evaluate these Cowley's diagrams is the calculation of their force constants (FC) of various orders. Nowadays, it is possible to compute them completely \emph{ab intio} using density functional theory (DFT) \cite{kohn1,kohn2} by following one of two main approaches. 
The first method employs density functional perturbation theory (DFPT) \cite{baroni2001phonons}, and the FC are expressed directly in terms of the electron density and its derivatives\cite{paulatto2013anharmonic}. In the second one, the finite displacement method (FDM) \cite{parlinski1997first, kresse1995ab}, the FC are extracted by performing a series of self-consistent calculations for different atomic displacement patterns and using the Hellman-Feynman theorem.
However, the main issue in this approach is that, the computations grow rapidly since a large number of configurations is needed for higher order FC. The calculation of the FC of the $m$-th order requires $(6Nn_b)^m$ configurations, where $n_b$ is the number of atoms in the lattice basis, and $N$ is the number of primitive cells in the supercell \cite{feng}. Although, this number can be greatly reduced by identifying equivalent displacement patterns using crystal symmetries \cite{stokes,feng}, it is still quite large, especially for the fourth-order FC. 
In addition to Cowley's contributions to the anharmonicity, the effects derived from isotope-disorder scatterings have been shown to play an important role at low temperature \cite{fugallo2018infrared}. The formulation to deal with this isotopic scattering can also be described in the second-order perturbation theory \cite{tamura1983isotope}. In this framework, the unperturbed Hamiltonian, $H_0$, for the crystal is defined in the harmonic approximation with the atomic mass replaced by an averaged mass that depends on the number of unit cells in the crystal and the fraction of a given atomic isotope. Then, the first perturbation, $H_I$, which depends on the deviation of the isotope mass from the average mass, has its main contribution from the second-order term shown in Fig.~\ref{phon_diag}(d). The first-order and other higher-order diagrams derived from $H_I$ have either zero contribution (due to the average mass definition) or are not significant in comparison to the second-order term \cite{tamura1983isotope}. Therefore, in order to study the most substantial anharmonic effects in MgS and CaS, we use the FDM including only three-phonon processes and the isotopic scattering shown in Fig.~\ref{phon_diag}.

The rest of this article is structured as follows. In Section II we summarize the necessary elements to model the infrared response as well as the means to calculate them. We present our results on optical properties in Section III and compare them to available theoretical and experimental findings within the harmonic limit and highlight the anharmonic contributions. Finally, in Section IV we draw our conclusions and argue for the applicability of the approach used in this work to model infrared optical properties of other materials. 
\section{\label{sec:level1}Methodology and Computational details}
Our main goal is to calculate optical response of MgS and CaS in the infrared region and to this end, we use the generalized Lorentz model. In this scheme the dielectric function $\epsilon$ as a function of frequency $\omega$ is given by:
\begin{equation}
    \epsilon(\omega) = \epsilon_{\infty} + \sum_j\frac{S_j}{\omega^2_{j,0} - \omega^2 + 2\omega\Pi_j(\omega)}, \label{dielecfun}
\end{equation}
where $\epsilon_{\infty}$ is the high-frequency dielectric constant, $s_j$ and $\omega_{j,0}$ are the strength and frequency of the $j$-th oscillator, respectively. The oscillator strength is related to the Born effective charges $Z$, as $S_j = 4\pi Z^2\slash (v\mu_j \omega_{j,0}^2)$, where  $v$ is the unit-cell volume and $\mu$ is the reduced mass of the oscillator \cite{fugallo2018infrared}. Lastly, $\Pi_j(\omega)$ is the self-energy of the $j$-th vibrational mode.

The self-energy can be  improved gradually according to Cowley's perturbative approach \cite{cowley1963lattice}. In our work, we restrict its expansion to the following form:
\begin{equation}
    \Pi_{tot}(\omega) = \Pi_{3ph}(\omega) + \Pi_{isot}(\omega) \label{selfen},
\end{equation}
where $\Pi_{3ph}$ is the anharmonic contribution to the total self-energy due to three-phonon scattering processes, and $\Pi_{isot}$ is the isotope-disorder induced scattering (Figs.~\ref{phon_diag}(a)-(b) and \ref{phon_diag}(d)). The three-phonon part of the self-energy consists of two terms known as the ``bubble'' and ``tadpole'' diagrams. The bubble term has both real and imaginary parts whereas the tadpole has only the real part. Thus, the finite lifetime of the phonons due to the phonon-phonon interaction is solely determined by the imaginary part of the bubble (B) diagram, Im$\left[\Pi^B_{3ph}(\omega)\right]$, which is the only term considered in our calculation. We note that in general, the bubble diagram is not diagonal in branch indices $jj^{\prime}$, however, for the purpose of calculating optical properties, it is sufficient to consider only the diagonal terms\cite{maradudin1962scattering}. The second term in Eq. (\ref{selfen}) also contributes to the broadening of the harmonic frequencies. $\Pi_{isot}(\omega)$ is mostly visible at low frequencies and temperatures and, as we show in the next section for our sulfide systems, it becomes smeared out by the phonon-phonon interaction at high temperatures \cite{tamura1983isotope,fugallo2018infrared}.

In this study, we mainly seek to model the reflectivity, which is characterized by a frequency band of high reflectivity, i.e., the reststrahlen band \cite{calandrini2021limits}. The top and bottom of this band are defined by the transverse ($\omega_{TO}$) and longitudinal ($\omega_{LO}$) optical modes, respectively. These frequencies can be calculated from the harmonic approximation alone by employing the non-analytical correction (NAC) at {\it almost} zero $q$-vector \cite{gonze1997dynamical}. It also means that the summation in Eq. (\ref{dielecfun}) is reduced to only those modes that fall within the reststrahlen band, i.e., the transverse optical (TO) mode.

All our calculations are performed using DFT and the projector-augmented plane wave method (PAW) \cite{blochl} as implemented in the VASP  code (version 5.4.4) \cite{kresse1996efficiency,kresse1996efficient}. The valence configurations are 3\emph{p$^6$}4\emph{s$^2$} for Ca, 2\emph{p$^6$}3\emph{s$^2$} for Mg, and 3\emph{s$^2$}3\emph{p$^4$} for S. The exchange-correlation (XC) term in the effective Kohn-Sham potential is approximated according to  the Perdew-Burke-Ernzerhof parameterization for solids (PBEsol)\cite{PBEsol} of the generalized gradient approximation (GGA)\cite{ggaoriginal}. Both CaS and MgS crystallize into rock-salt (RS) cubic structure. We use their conventional  unit-cells  with 8 atoms. Integration in the Brillouin zone is done on a $\Gamma$-centered grid of uniformly distributed $k$-points with a spacing of $2\pi\times0.3$~ \AA$^{-1}$. The selected plane-wave kinetic energy cutoff is 500~eV and convergence of our structural optimizations is assumed when the total energy changes are less than $10^{-8}$~eV and the forces on each atom smaller than $10^{-3}$~eV/\AA. We use fully relaxed crystal structures (in unit-cell shape and ionic degrees of freedom)  as the underlying unit-cells for the generation of atomic displacements.

The displacement configurations are generated in accordance with the crystal symmetry as implemented in the \emph{Phonopy}\cite{phonopy} code. We use $2 \times 2 \times 2$ supercells, which contain 64 atoms, and to obtain second-order FC, two displacement configurations are sufficient. Then, we  calculate the harmonic eigenfrequencies with NAC to estimate the size of the reststrahlen band. To compute the third-order FC necessary to calculate the self-energy, i.e., $\Pi^B_{3ph}(\omega)$ \cite{maradudin1962scattering}, we also use the FDM method as  implemented in the \emph{Phono3py}\cite{phono3py} code. Unlike for the second-order FC, for the third-order case every pair of atoms in the supercell needs to be displaced, but in this instance, the crystal symmetry once again helps reducing the overall number of required configurations. The phonon interaction distance (i.e., the distance between the displaced atoms in the configurations) is chosen to be 7~\AA\space and 8~\AA\space for CaS and MgS, respectively. These parameters allow us to achieve the largest number of displacement patterns possible, that is, 146 unique displacement configurations for both systems. This condition of maximal displacements is necessary to converge the self-energy. Finally, we compute the self-energy on a uniform $12 \times 12\times 12$ $q$-point grid, resulting in the overall number of 720 $q$-points in the first Brillouin zone. 
\section{Results: Optical properties}
\subsection{Harmonic approximation limit}
\begin{figure}
\includegraphics[width=1\linewidth]{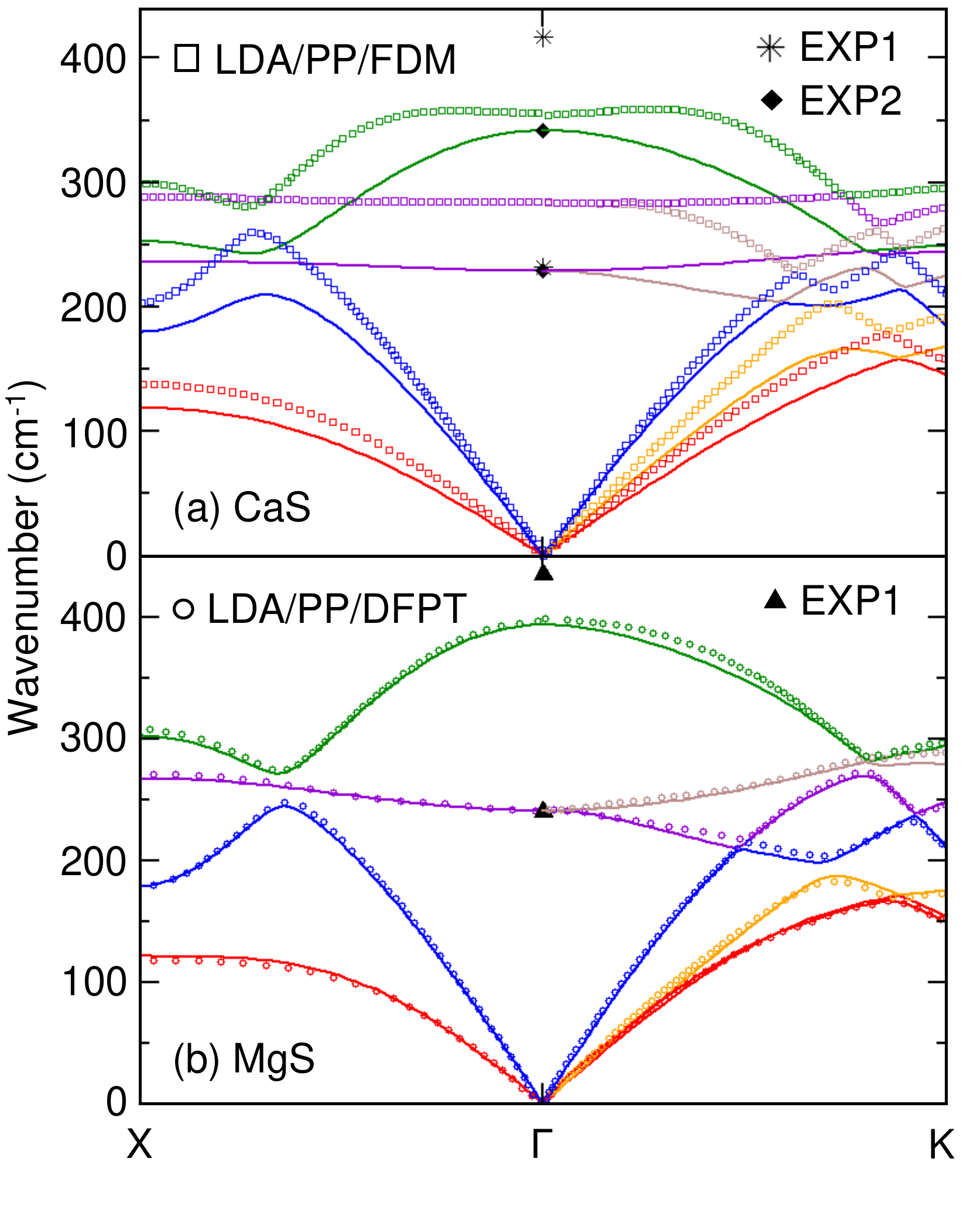}
\caption{\label{bands}Our simulated phonon bands within the harmonic approximation are shown in solid lines for (a) CaS and (b) MgS. Different colors correspond to individual normal modes of motion: three acoustic bands (red, orange, and blue), and three optical bands (violet, brown, and green are the two TO and one LO mode, respectively). Hollow squares \cite{bayrakci} (CaS) and hollow circles (MgS) \cite{duman2006first} represent data taken from previous modeling studies. Experimental values shown in solid symbols for the TO and LO modes are denoted by EXP1\cite{hofmeister2003absorption} and EXP2\cite{kaneko1982optical}.}
\end{figure}

\begin{table*}[t!]
\caption{\label{table1}Lattice parameter ($a$), transverse and longitudinal optical modes ($\omega_{TO}$ and $\omega_{LO}$), Born effective charge ($Z$) and high frequency dielectric constant ($\epsilon_{\infty}$) calculated in this work [*]. We compare to other theoretical and experimental data available in the literature: $\dagger$-Ref.~[\onlinecite{kaneko1982optical}], $\flat$-Ref.~[\onlinecite{hofmeister2003absorption}], $\ddagger$-Ref.~[\onlinecite{peiris1994compression}], $\star$-Ref.~[\onlinecite{bayrakci}], $\diamond$-Ref.~[\onlinecite{duman2006first}], $\bullet$-Ref.~[\onlinecite{straub1989self}],  $\triangleleft$-Ref.~[\onlinecite{boswarva1970semiempirical}], $\triangleright$-Ref.~[\onlinecite{pring_1998}]. Note that in the Ref.~[\onlinecite{kaneko1982optical}], the so-called Szigeti effective charge, $Z_S$, is given instead of $Z$. The Born effective charge we report here can be obtained using the relationship \cite{szigeti1949polarisability,szigetinew} $Z = \frac{(\epsilon_{\infty}+2)}{3}Z_S$. The high-frequency dielectric constant $\epsilon_{\infty}$ used in Ref.~[\onlinecite{kaneko1982optical}] and  Ref.~[\onlinecite{bayrakci}] was obtained from a semiempirical (SE) model in Ref.~[\onlinecite{boswarva1970semiempirical}]. PP: Pseudopotential.} 
\begin{ruledtabular}
\begin{tabular}{lllcccc}
System &Method  & $a$(RS)  & $\omega_{LO}$  & $\omega_{TO}$& $Z_{\text{Ca,Mg}}$&$\epsilon_{\infty}$  \\
         & &(\AA) & (cm$^{-1}$) & (cm$^{-1}$) & ($e$) \\
\colrule
CaS & PBEsol/FDM* & 5.633*   &341.55*  &228.69* &2.350* & 5.208* \\
    & LDA/PP/FDM$^\star$,SE$^\triangleleft$    &5.670$^\star$    &354.00$^\star$  & 284.00$^\star$ & 1.802$^\bullet$&4.150$^\triangleleft$ \\
    & EXP1$^\flat$   &    &417.00$^\flat$  &232.00$^\flat$  & &4.580$^\triangleright$  \\
    & EXP2$^\dagger$      &5.697$^\dagger$   & 342.00$^\dagger$ &229.00$^\dagger$  &2.111$^\dagger$&4.150$^\triangleleft$   \\
MgS & PBEsol/FDM* &5.180*   &393.00*  &240.60*  &2.314* & 5.542* \\
    & LDA/PP/DFPT$^\diamond$    &5.180$^\diamond$&397.00$^\diamond$  &241.00$^\diamond$  &2.350$^\diamond$&5.660$^\diamond$   \\
    & EXP1$^\flat$      &5.200$^\ddagger$    &435.00$^\flat$  &240.00$^\flat$  & &4.800$^\triangleleft$  \\
\end{tabular}
\end{ruledtabular}

\end{table*}
Firstly, we analyze the dynamical properties of CaS and MgS within the harmonic approximation. As we discuss above, it is imperative to establish correctly the bounds of the reststrahlen band needed subsequently to calculate the reflectivity. In Table~\ref{table1}, we summarize our predicted values for the lattice parameter $a$, the transverse $\omega_{TO}$ and longitudinal $\omega_{LO}$  optical modes, the Born effective charge $Z$, and the high frequency dielectric constant $\epsilon_{\infty}$ for CaS and MgS. Their computed phonon band spectra are shown in Fig.~\ref{bands}(a) and Fig.~\ref{bands}(b), respectively.\\ 
Our calculated phonon band structure for CaS is in good qualitative agreement with a previous LDA/Pseudopotential/FDM study \cite{bayrakci} using the SIESTA implementation\cite{siesta} as shown in Fig.~\ref{bands}(a). The main quantitative differences are due to lower frequencies predicted in our work for all phonon bands. This incosistency in frequency magnitudes could be partially attributed to our slightly smaller lattice parameter $a_{\text{CaS}}=5.633$~\AA \ with respect to $a_{\text{CaS}}=5.670$~\AA \ of Ref.~[\onlinecite{bayrakci}], which is slightly in better agreement with single crystal measurements \cite{kaneko1982optical} of $a_{\text{CaS}}=5.697$~\AA. However, our phonon band structure seems to be more consistent with experimental data\cite{kaneko1982optical,hofmeister2003absorption}, as our predicted LO/TO splitting of 112.86~cm$^{-1}$ agrees more with reported values of 113~cm$^{-1}$ by Ref.~[\onlinecite{kaneko1982optical}]. From Table~\ref{table1}, we notice that the two experimental studies find nearly the same value for $\omega_{TO}$, but they differ in the magnitude of $\omega_{LO}$. An apparent reason for this discrepancy could be traced back to the method employed to extract $\omega_{TO}$ and $\omega_{LO}$ from experiments. Both studies use a single oscillator model in their dispersion analysis, and adopt an expression virtually identical to Eq. (\ref{selfen}), but they treat the self-energy as frequency independent, giving it the role of a {\it damping constant}. In EXP1 \cite{kaneko1982optical} (in combination with $\epsilon_{\infty}$ taken from Ref.~[\onlinecite{pring_1998}], the real and imaginary parts of the dielectric function, Re($\epsilon$) and Im($\epsilon$), are constructed from reflectivity data, then $\omega_{TO}$ and $\omega_{LO}$ are taken as the maxima of Re($\epsilon$) and Inv[Im$(\epsilon)$], respectively. 
On the other hand, in EXP2 \cite{hofmeister2003absorption}  the TO and LO frequencies are treated as adjustable parameters to parametrize the reflectivity and fit the data. Thus, although the resolution of $\omega_{TO}$ seems to be independent of the method, the latter seems to be more in agreement with our computation.
Nevertheless, the calculated LO/TO splitting of 70~cm$^{-1}$ by Ref.~[\onlinecite{bayrakci}] is underestimated by at least 38\% with respect to both measurements. This underestimation could be linked to their Born charge of $Z_{\text{Ca}}=1.802e$ and high-frequency dielectric constant $\epsilon_{\infty}=4.150$ taken from Ref.~[\onlinecite{straub1989self}] and Ref.~[\onlinecite{boswarva1970semiempirical}], respectively. 
In contrast, our Born charge of $Z_{\text{Ca}}=2.350e$, calculated directly from our relaxed CaS structure, is much closer to the experimental value of $2.111e$. Here, we observe clearly that a self-consistency among the required elements entering in the calculation of  $\omega_{TO}$ and $\omega_{LO}$ is highly desirable, in particular, the LO/TO splitting is rather sensitive to the Born effective charge, as it enters the NAC as $Z^2$. An accurate determination of the TO mode is rather significant for the modeling of the reflectivity as it controls the edge of the reststrahlen band \cite{calandrini2021limits}. 
 
In the case of MgS, our predicted lattice parameter $a_{\text{MgS}}=5.180$~\AA \ (Table~\ref{table1}) and phonon band spectrum, (Fig.~\ref{bands}(b)) are in excellent agreement with an earlier LDA/Pseudopotential/DFPT study \cite{duman2006first} as implemented in the quantum espresso code\cite{quantumespr}. Our Born effective charge $Z_{\text{Mg}}=2.31e$ and high-frequency dielectric constant $\epsilon_{\infty}=4.150$ are also almost identical to the values found in Ref.~[\onlinecite{duman2006first}] of $Z_{\text{Mg}}=2.314e$ and $\epsilon_{\infty}=4.150$. These findings further support the idea that the main reason for the significant difference between our results and Ref.~[\onlinecite{bayrakci}] in the case of CaS stemmed from the sizeable difference in the values of $Z_{\text{Ca}}$ and $\epsilon_{\infty}$. However, our LO/TO splitting of 152.4~cm$^{-1}$, and the other {\it ab initio} value of 156~cm$^{-1}$ are both, about 20\% smaller than absorption spectra data \cite{hofmeister2003absorption}  reporting a LO/TO splitting of 195~cm$^{-1}$. But, as discussed previously for the CaS system, the method used by Ref.~[\onlinecite{hofmeister2003absorption}] to determine $\omega_{TO}$ and $\omega_{LO}$ seems to give an overestimated value for $\omega_{LO}$ (Table \ref{table1}), which could be the main source of divergence between the calculated and the experimental values. 
\subsection{Anharmonic effects}
We now turn our attention to the anharmonic effects and their role in the optical properties of CaS and MgS. As explained in the previous section, the only contributions to the self-energy that we consider are the three-phonon $\Pi_{3ph}$ scattering processes and the isotopic disorder $\Pi_{isot}$.   

\begin{figure}
\includegraphics[width=1\linewidth]{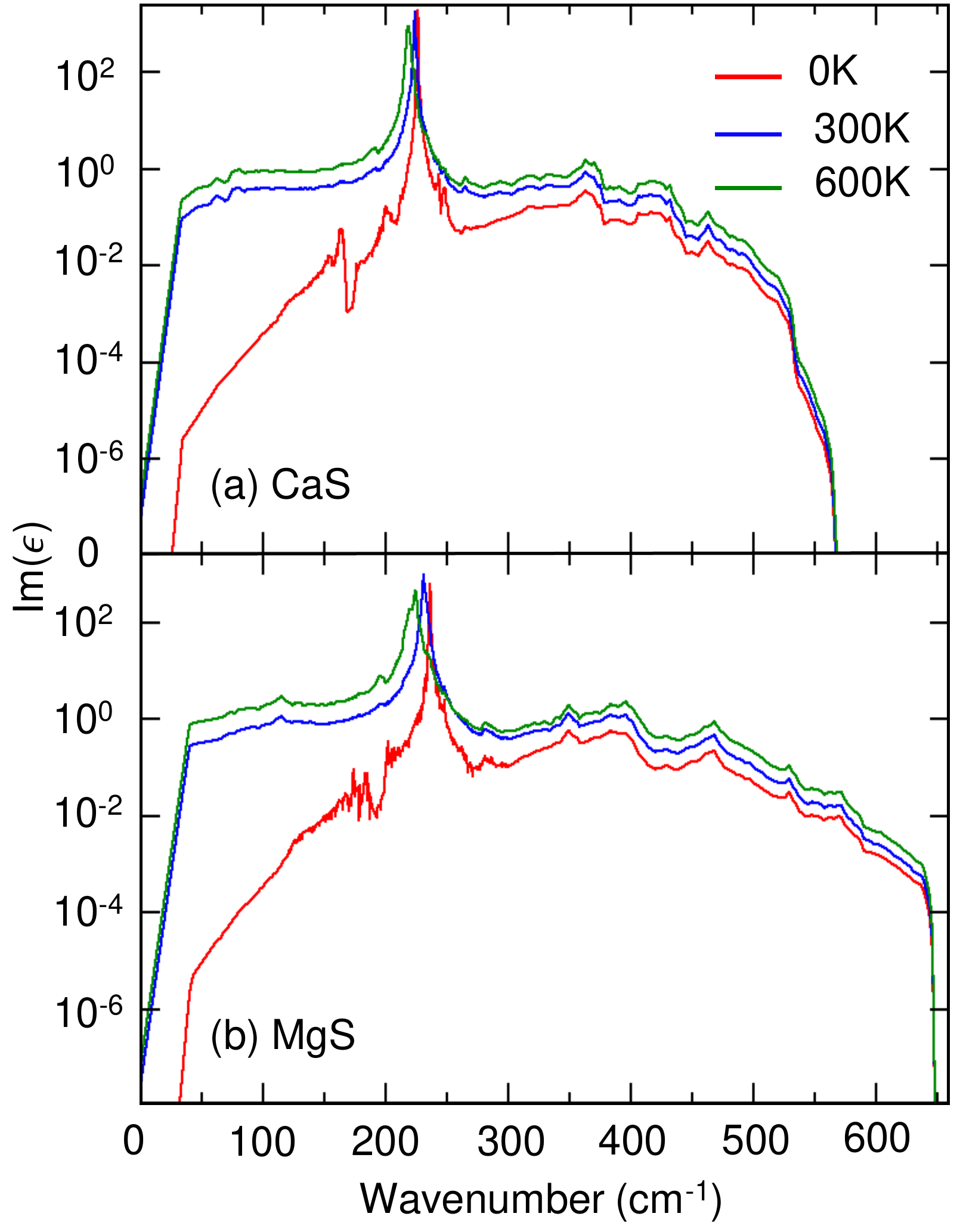}%
\caption{\label{eps} Calculated imaginary part of the dielectric function, Im[$\epsilon$], for (a) CaS and (b) MgS at three different temperatures. The isotope-disorder scattering effects are responsible for the oscillatory behavior in the region 150-220~cm$^{-1}$ and are only visible at sufficiently low temperature. The three-phonon processes prevail over $\Pi_{isot}$ at higher temperatures.}
\end{figure}

We notice that  the $\Pi_{isot}$ contribution is only prominent at very low temperature, and its effect can be seen clearly in the imaginary part of the dielectric function, Im$(\epsilon)$, as shown in Fig.~\ref{eps}. Three-phonon processes are essentially absent at low temperature and low wavenumbers, thus the isotopic disorder is the leading term in this region. But as temperature increases, the  three-phonon processes become dominant and smear out the isotope-disorder part. In addition, one can also observe the Im$(\epsilon)$ peak slightly shifting towards lower wavenumbers. Higher-order scattering processes (e.g., four-phonon contributions) are expected not to have a drastic effect on the computed Im$(\epsilon)$ of MgS and CaS. This assumption is in line with modeling results for MgO \cite{fugallo2018infrared}, as it was demonstrated that the four-phonon processes start being substantial at relatively high wavenumbers when the three-phonon contributions vanish. \\
%
\begin{figure}
\includegraphics[width=1\linewidth]{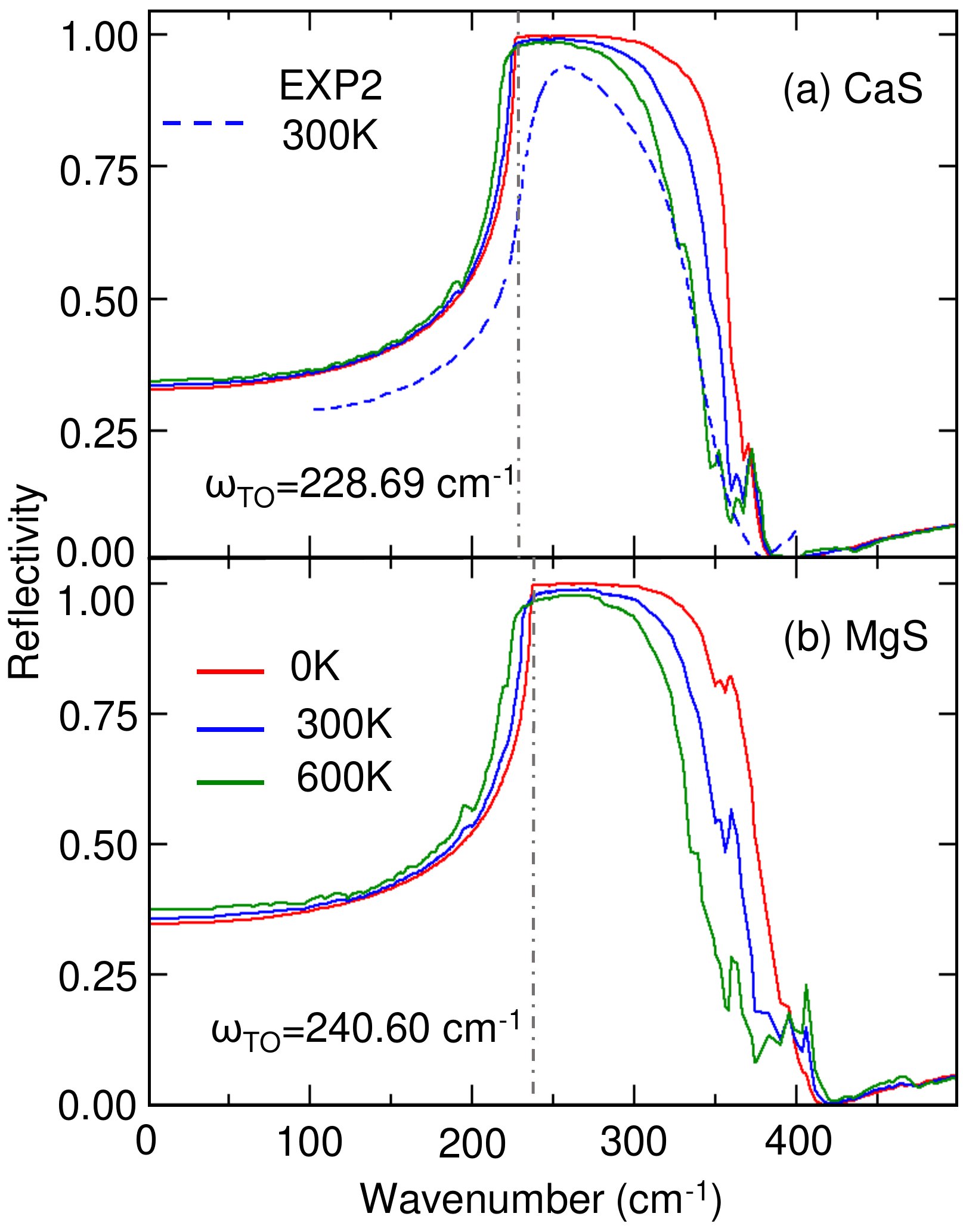}
\caption{\label{refl} Predicted reflectivity (solid lines) for (a) CaS and (b) MgS at three different temperatures. For CaS, we compare our results to an experimental trend \cite{kaneko1982optical} (dashed line) at 300~K. The dotted-dashed gray lines indicate the position of our predicted $\omega_{TO}$ modes. 
}
\end{figure}
Our calculated reflectivity at three different temperatures for CaS and MgS is shown in Fig.~\ref{refl}. Here, we can immediately see the importance of the precise prediction of the $\omega_{TO}$ value, as it defines the reflectivity maximum. The width of the plateau at low temperature is mostly determined by the high-frequency dielectric constant $\epsilon_{\infty}$  and the Born effective charge $Z$. As temperature increases,  both edges of the plateau smear out, although the change is much more obvious at the higher-wavenumber end. This effect is due to the anharmonic interactions, which become more notorious with increasing temperature. For MgS, Fig.~\ref{refl}(b), we observe characteristic shoulders between  $\sim$325~cm$^{-1}$ and $\sim$425~cm$^{-1}$, these features appear in MgO\cite{fugallo2018infrared,calandrini2021limits} as well. We also notice that for MgS, these shoulders show a somewhat spiky structure. For CaS, these shoulders are more prominent as temperature increases, that is, for $\ge300$~K. The only available experimental report is on the reflectivity of CaS \cite{kaneko1982optical} as shown in Fig.~\ref{refl}(a). If we compare it to our reflectivity computed at 300 K, we observe a good qualitative agreement with some systematic quantitative deviation. Our predicted position of the TO peak is only 0.14\% lower than the measured one at 229~cm$^{-1}$. In general, we notice that the experimental reflectivity values are smaller  than our simulation results in the region considered. It is highly unlikely that the inclusion of additional anharmonic terms in the Cowley's expansion of  Eq. (\ref{selfen}) would reduce the overall magnitude of our predicted reflectivity. 
Namely, we think that the main cause for the overestimation of the reflectivity has its roots in the calculated force constants, and consequently, the estimated strength of the anharmonic effects. Perhaps, larger supercells could improve the overall FC values because they would allow for larger interaction distances between phonons, however, the computational cost would increase significantly. The other peculiar feature of our predicted reflectivity curves for  CaS and MgS is the presence of several peaks in the high-wavenumber shoulders. Such structures are not observed in the experimental study on CaS\cite{kaneko1982optical}, where all possible anharmonic effects are, of course, present. It is conceivable that the inclusion of higher-order phonon processes could remove these spiky structures in that region as it is shown in the case of MgO\cite{fugallo2018infrared}, in which both three- and four-phonon processes were included and a smooth computed reflectivity is achieved. Therefore, we speculate that at the level of our simulations, both CaS and MgS seem to be more anharmonic in the sense that, more anharmonic terms are needed to be included in order to reach a better agreement with the experimentally observed reflectivity, but such studies are outside of our current resources and are left for future explorations.

At this point, it is worth mentioning that the experimental results used for comparison in our previous discussion are from single crystal measurements. However, there are experiments in which, powdered  (pressed pellets) samples are studied also with the infrared spectroscopy technique. This experimental approach is particularly relevant in planetary investigations, as {\it dust} is likely to be studied directly on the planet's surface thanks to spectrometers in missions or in simulated conditions in the laboratory \cite{VARATHARAJAN2019127}. However, the use of pressed pellets leads to scattering, and diffused reflection or refraction, resulting in an overall weaker reflectivity measurement \cite{johnson2020infrared}. It has been also observed that, sometimes, reflectivity spectra from powdered samples retain qualitative features of the reflectivity of single crystals, that is, the measured peaks do not shift in wavenumber but only decrease in magnitude. \cite{long1993optical}. In such scenario, we expect the harmonic modeling using the generalized Lorentz model could still be applicable, but adjusted to measurements from sulfide powders, by modifying the number of oscillators and the damping coefficient. Thus, the self-energy  (which plays the role of the damping factor) in Eq. (\ref{selfen}), would be then dominated not by anharmonic contributions, but by scatterings, grain sizes, etc., i.e., processes specifically differentiating between single crystal and pressed pellets.

In recent years, with the objective of building a spectral library of sulfide minerals to support investigations of Mercury's surface chemistry, a study \cite{VARATHARAJAN2019127} reported reflectance ($R$) measurements of synthetic CaS and MgS powders in the range from far infrared (FIR) until visible (VIS), at day Mercury's surface temperature ($\sim$773~K during the day). Reflectance is related to reflectivity in that the latter is the reflectance of a semi-infinite slab \cite{fugallo2018infrared}. Thus a direct and straight comparison between their reflectance magnitudes is not possible, nonetheless, the position of their maxima, which should coincide, can still be investigated. The measured $R$ maximum in the FIR region for MgS (sample denoted ``MgS-2'' in Ref.~[\onlinecite{VARATHARAJAN2019127}] occurs at the wavelength of 40~$\mu$m, equivalent to a wavenumber of 250~cm$^{-1}$. In comparison, we predict the maximum of the reflectivity at 240~cm$^{-1}$ and 300~K. Finally, for CaS, the reflectance measurements\cite{VARATHARAJAN2019127} position its peak at a wavelength of about 8~$\mu$m, that is, a wavenumber of about 555~cm$^{-1}$, this value is in drastic disagreement with our calculated position for the reflectivity peak at 228.69~cm$^{-1}$ and the single crystal measurements \cite{kaneko1982optical,hofmeister2003absorption} of 229 and 232~cm$^{-1}$. This sharp difference in the position of the maxima from reflectance and reflectivity could be due to the use of CaS powder in the case of the reflectance experiment \cite{VARATHARAJAN2019127}, however, further experiments would be necessary to clarify the discrepancy.

\section{Conclusions}
In summary, we have conducted a full first-principles study of the reflectivity of CaS and MgS in the infrared region. We have considered both harmonic and anharmonic contributions. Within the harmonic limit, we have demonstrated that it is crucially important to employ the non-analytic correction to obtain the correct LO/TO splitting. An accurate determination of the TO and LO peaks is highly desirable as they provide the boundaries of the reststrahlen band, i.e.,  the low- and high-frequency edges of the maximum reflectivity. We have also shown the effects of the anharmonicity by considering three-phonon scatterings and isotope-disorder processes at the lowest perturbation level. The anharmonic terms' main influence occurred in the smear of the reflectivity spectra's edges, being more prominent in the higher-wavenumber region.

Although we have not included higher-order anharmonic terms, e.g., four-phonon scatterings, we think that these processes would be only noticeable in the high-wavenumber region of our study, outside of the maximum reflectivity peak, as it has been shown to be the case for MgO \cite{fugallo2018infrared}. However, these higher-order anharmonic terms could eliminate the spiky structures in our predicted reflectivities. Finally, it might be worthwhile to try to incorporate the anharmonic terms into the modeling of the polycrystalline and powder samples. Although, in this case, we expect them to play a secondary role since the self-energy, i.e, the damping constant would be dominated by the processes that are characteristic of powder pellets such as diffused reflection and refraction.
\begin{acknowledgments}
The authors gratefully acknowledge the Gauss Centre for Supercomputing e.V. (www.gauss-centre.eu) for funding this project by providing computing time through the John von Neumann Institute for Computing (NIC) on the GCS Supercomputer JUWELS at J\"ulich Supercomputing Centre (JSC) under project {\bf geopressphon}. AC and MNV were supported by the Helmholtz Association through {\it funding of first-time professorial appointments of excellent women scientists (W2/W3)}
\end{acknowledgments}
\bibliographystyle{apsrev4-1}
\bibliography{bibliography}
\end{document}